\documentclass[namedreferences,second paper]{SolarPhysics}
\usepackage[optionalrh,solaenum]{spr-sola-addons}
\usepackage{latexsym}
\usepackage{wrapfig}
\usepackage{epsfig}
\usepackage{graphicx}
\usepackage{subfig}
\usepackage{color}           
\usepackage{url}             
\usepackage[english]{babel}
\newcommand{\md}{\mathrm{d}}

\begin{document}

\begin{article}

\begin{opening}
\title{\center{The Effect of Variable Background on Oscillating Hot Coronal Loop due to Thermal Conduction}}

\author{K.S.~\surname{Al-Ghafri} and R.~\surname{Erd\'{e}lyi}}

\runningauthor{K.S.~\surname{Al-Ghafri} and R.~\surname{Erd\'{e}lyi}}
\runningtitle{The Effect of Variable Background on Oscillating Hot Coronal Loop due to Thermal Conduction}

\institute{Solar Physics and Space Plasma Research Centre (SP$^2$RC), University of Sheffield,\\ Hicks Building, Hounsfield Road, Sheffield S3 7RH, UK \\
e-mail: \url{app08ksa@sheffield.ac.uk}\\
e-mail: \url{robertus@sheffield.ac.uk}}
\begin{abstract}
We investigate the effect of a variable, {\it i.e.} time-dependent, background on the standing acoustic ({\it i.e.} longitudinal) modes generated in a hot coronal loop. A theoretical model of 1D geometry describing the coronal loop is applied. The background temperature is allowed to change as a function of time and undergoes an exponential decay with characteristic cooling times typical for coronal loops. The magnetic field is assumed to be uniform. Thermal conduction is the dominant mechanism of cooling the hot background plasma in the presence of an unspecified thermodynamic source that maintains the initial equilibrium. The influence of the rapidly cooling background plasma on the behaviour of standing acoustic (longitudinal) waves is investigated analytically. The temporally evolving dispersion relation and wave amplitude are derived by using the WKB theory. An analytic solution for the time-dependent amplitude that describes the influence of thermal conduction on the standing longitudinal (acoustic) wave is obtained by exploiting the properties of Sturm-Liouville problems. Next, numerical evaluations further illustrate the behaviour of the standing acoustic waves in a system with variable, time dependent background. The results are applied to a number of detected loop oscillations. We find a remarkable agreement between the theoretical predictions and the observations. The cooling of the background plasma due to thermal conduction is found to cause a strong damping for the slow standing magneto-acoustic waves in hot coronal loops in general. Further to this, the increase in the value of thermal conductivity leads to a strong decay in the amplitude of the longitudinal standing slow MHD waves.
\end{abstract}
\end{opening}

\section{Introduction}
Recent consecutive solar observations by high-resolution imaging space telescopes and spectrometers have shown that the solar atmosphere is dynamic in nature and is composed of numerous magnetic structures of which coronal loops are of the centre of focus in this paper. It has been confirmed that these complex structures of the solar corona can support a wide range of magnetohydrodynamic (MHD) waves which are natural carriers of energy and may have the key to solve the problem of solar coronal heating (see, the recent review by {\it e.g.} \opencite{Taroyan09}; \opencite{McLaughlin11}). In particular, one MHD wave mode present in coronal loops has become the centre of attention: the slow magneto-acoustic mode. The slow (propagating and standing) MHD waves are extensively supported by the solar coronal structures and are observed to be rapidly damped \cite{Wang03a,Wang03b,Moortel09,Wang11}.

Propagating intensity disturbances were first detected by the Ultraviolet Coronagraph Spectrometer onboard the {\it Solar Heliospheric Observatory} (SOHO/UVCS) along coronal plumes \cite{Ofman97, Ofman99,Ofman001a}, and were identified as slow magneto-acoustic waves \cite{Ofman99}. Subsequently, similar intensity disturbances were observed in coronal loops by the {\it Transition Region and Coronal Explorer} (TRACE) \cite{Nightingale99,Schrijver99,Moortel00,McEwan06} and the EUV Imaging Telescope onboard the {\it Solar Heliospheric Observatory} (SOHO/EIT) \cite{Berghmans99}. \inlinecite{Nakariakov00} found that the slow MHD wave evolution is affected by dissipation and gravitational stratification. \inlinecite{ErdelyiTaroyan08} and \inlinecite{Wang09} have detected longitudinally propagating slow MHD waves with around five-minute period in the transition region and five coronal lines at the footpoint of a coronal loop by {\it{Hinode}}/EIS. The source of these oscillations are suggested to be the leakage of the {\it p} modes from photosphere region through the chromosphere and transition region into the corona \cite{Pontieu05}.

Oscillations interpreted as longitudinal standing (slow) magneto-acoustic waves have been observed in hot ($T>6$~MK) active region loops by the Solar
Ultraviolet Measurement of Emitted Radiation (SUMER) spectrometer on board SOHO \cite{Wang02,Wang03b,Taroyan07}. These oscillations have periods in the range of $8.6$ to $32.3$ minutes with decay times of $3.1-42.3$ minutes and amplitudes between 12 and 353 km s$^{-1}$ \cite{Wang05}. Moreover, \inlinecite{Mariska05} have reported Doppler shift oscillations during solar flares with Yohkoh in a high temperature region reaching $12-14$~MK. These oscillations are interpreted in terms of the standing-slow mode MHD waves \cite{Mariska06}. Evidence for the standing slow mode can be underpinned from the phase relationship between velocity and intensity where a quarter-period phase difference is a characteristic of the standing waves while the propagating waves exhibit an in-phase relationship. Therefore, an approximate quarter-period delay of the intensity variations behind the Doppler shift strongly support that the oscillations observed by SUMER are slow standing modes. The observed oscillations in coronal loops indicate that the standing slow modes are likely triggered by micro-flares which are produced by impulsive heating \cite{Mendoza02}. In a recent work, \inlinecite{Taroyan08b} found that longitudinal standing waves can be excited in cooler EUV loops under the effect of all important mechanisms such as gravitational and thermal stratification, losses, etc. According to this study, hot loops are not only the origin of the standing waves but these waves can also be formed in cooler loops.

Nowadays, the damping of slow magneto-acoustic waves have become a subject of remarkable observational and theoretical attention due to a possibility of damping timescale of the waves on disclosing the physical processes that dominate the energy of the coronal loop in which they are detected. The majority of theoretical and numerical studies on damping of propagating and standing slow MHD waves show that the understanding of dominant mechanisms of rapid damping can be captured from 1D linear \cite{Sigalotti07} or nonlinear model \cite{Wang11}. For instance, \inlinecite{Ofmanwang} and \inlinecite{Mendoza04} found that the standing MHD waves are strongly damped because of thermal conduction in nonlinear model whereas using a linear MHD model \inlinecite{Pandey06} indicated that the individual influence of thermal conduction or viscosity is not enough to account for the observed damping. In a static 1D isothermal medium, \inlinecite{Moortel03} investigated the behaviour of both propagating and standing slow MHD waves in the presence of thermal conduction and compressive viscosity. They found that  thermal conduction is the dominant damping mechanism of thermal and magneto--acoustic waves in coronal loops. It is also shown that thermal mode is damped in the form of standing waves. Moreover, \inlinecite{Morton09d} have shown that the radiative cooling causes a damping of the slow mode by up to 60$\%$ within characteristic lifetimes.

\inlinecite{Mendoza04} studied the influence of gravitational stratification on damping of standing MHD waves in hot coronal loops and found that enhanced nonlinear viscous dissipation due to gravity may reduce the damping times by about $10-20\%$ compared to the unstratified loops. In contrast to most recent work, \inlinecite{Sigalotti07} found that thermal conduction can only be accounted as damping mechanism when the compressive viscosity is added to the model. \inlinecite{Bradshaw08} reported that the radiation due to a non-equilibrium ionisation balance could cause by up to $10\%$ reduction of wave-damping timescale in comparison to the equilibrium case. \inlinecite{Verwichte08} showed that shock dissipation at large amplitudes gives rise to enhancement of the damping rate which is up to $50\%$ larger than given by thermal conduction alone. In non-isothermal, hot, gravitationally stratified coronal loops, \inlinecite{Erdelyi08a} investigated the damping of standing slow (longitudinal) waves and found that the decay time of waves decreases with the increase of the initial temperature. Further to this, they derived a relation between the damping time and the parameter determining the apex temperature in the form of second-order scaling polynomial. In all these earlier studies the background (equilibrium) plasma was considered time-independent. This is certainly a working model as long as the background plasma changes occur on timescales much longer than typical perturbation timescales. In a recent theoretical work by \inlinecite{Erdelyi11}, the cooling of the background plasma due to thermal conduction has been found to {\it decrease} the rate of damping of the propagating magneto--acoustic waves in a weakly stratified atmosphere where temporal changes in the slowly evolving equilibrium were considered.

In this study, we examine how the cooling background state affects the longitudinal {\it standing} waves in a uniform magnetised plasma. The additional complexity of a time-dependent background gives us a step along to model hot coronal loops. The cooling of the background plasma is assumed to be dominated by thermal conduction in the presence of an unspecified thermodynamic source where the plasma is assumed to be cooling in time with an exponential profile. The coefficient of thermal conductivity is presumed to be varying as a function of time to investigate how the variation of thermal conduction influences the behaviour of standing slow-mode waves. The geometry of coronal loop is described by a semi-circular shape. An analytical solution for the time-dependant amplitude of the standing slow MHD waves is derived using the WKB theory with the aid of properties of Sturm-Liouville problems. The results exhibit that thermal conduction causes a significant variation on the amplitude of the standing magneto--acoustic modes with time. The amplitude of waves are found to suffer damping due to the cooling of the background plasma by thermal conduction. The influence of the ratio change of the period of standing oscillation to the cooling time scale on the rate of damping of hot-loop oscillations seems to be highly efficient, where increasing the value of this ratio deceases the strength of damping  rapidly.
\section{The Model and Governing Equations}
Consider a uniformly magnetised plasma in which the temperature is changing as a function of time due to thermal conduction and the density is a constant. The magnetic field is assumed to be uniform and aligned the $z$-axes, {\it i.e.} $\mathbf{B}_0=B_0\mathbf{\hat z}$.  Therefore, the background state can be described as follows:

{
\begin{wrapfigure}{l}{0.45\textwidth}
\vspace{-8pt}
\includegraphics[width=0.45\textwidth,height=0.12\textheight]{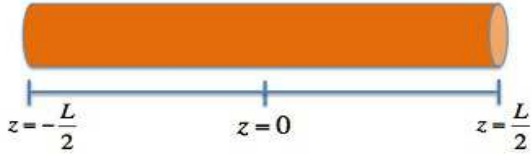}
\vspace{-25pt}
\caption{Coronal loop}
\vspace{30pt}
\end{wrapfigure}
\begin{eqnarray}
&&T_0=T_0(t),\\
&&p_0=p_0(t),\\
&&\rho_0=\textrm{const}.,\\
&&B_0=\textrm{const}.,\\
&&\epsilon=\frac{P}{\tau_c}\ll1.
\end{eqnarray}}\\
Here $T_0$, $p_0$, $\rho_0$ and $B_0$ are the background quantities identifying the temperature, pressure, density and magnetic field, respectively; and $P$ is the period of the loop oscillation and $\tau_c$ is the cooling time scale.
\\
\\
The governing MHD equations for the background plasma take the following form
\begin{eqnarray}
&&\frac{\partial{\rho}}{\partial{t}}+\nabla.(\rho\mathbf{v}) =0,\label{Eq:cont}\\
&&\rho\frac{\partial{\mathbf{v}}}{\partial{t}}+\rho(\mathbf{v}.\nabla)\mathbf{v}=-\nabla{p}+\frac{1}{\mu_0}(\nabla\times\mathbf{B})\times\mathbf{B},\\
&&\frac{R}{\tilde\mu}\rho^\gamma\left[\frac{\partial{}}{\partial{t}}\frac{T}{\rho^{\gamma-1}}
+(\mathbf{v}.\nabla)\frac{T}{\rho^{\gamma-1}}\right]=(\gamma-1)\nabla(\kappa_{\|}\nabla{T})+\mathcal{L} ,\\
&&{p}=\frac{R}{\tilde\mu}\rho{T},\label{Eq:gas-law}\\
&&\frac{\partial{\mathbf{B}}}{\partial{t}}=\nabla\times(\mathbf{v}\times\mathbf{B}),
\end{eqnarray}
where $\mathbf{v}$ is the flow velocity, $\mathbf{B}$ is the magnetic field, $g$ is the gravity, $\mu_0$ is the magnetic permeability of free space, $\gamma$ is the ratio of specific heats, $R$ is the gas constant, $\tilde{\mu}$ is the mean molecular weight, $T$ is the temperature, $\nabla(\kappa_{\|}\nabla{T})$ is the thermal conduction term where $\kappa_{\|}=\kappa_0T^{5/2}$, $\mathcal{L}$ is an unspecified thermodynamic source term, $\rho$ and $p$ are the plasma density and its pressure, respectively.

Assuming that there is no background flow and the background density is constant, so the previous equations determining the background plasma state reduce to
\begin{eqnarray}
&&v_0=0,\quad\rho_0=\textrm{const}.,\\
&&\nabla p_0=0,\label{Eq:back-motion}\\
&&p_0=\frac{R}{\tilde{\mu}}\rho_0T_0,\qquad {\it i.e.}\; p_0\sim T_0,\label{Eq:back-gas_law}\\
&&\frac{R}{\tilde\mu}\rho_0\frac{\md{T_0}}{\md{t}}
=\mathcal{L},\label{Eq:back-energy}
\end{eqnarray}
where the 0 index denotes background quantity.
The observed cooling of coronal loops has been shown to be well-approximated by exponential profile for radiative-cooling loops (\opencite{Aschwanden08}; \opencite{Morton09b}, \citeyear{Morton09c}). More recently,  \inlinecite{Erdelyi11} found that the temperature decreases exponentially in time  due to thermal conduction in hot coronal loops. As a result, we will assume here that the cooling temperature profile of coronal loops has the form
\begin{equation}
T_0=T_{0i}\exp(-\frac{t}{\tau_c}),\label{cooling_time}
\end{equation}
where $T_{0i}$ is the initial temperature at $t=0$.

Let us perturb the background state, where all variables can be written in the form
$$
f(z,t)=f_0(t)+f_1(z,t),
$$
where the subscript 0 refers to the equilibrium quantities and the subscript 1 indicates the perturbed quantities. In this study we consider longitudinal motions only, {\it i.e.} $\mathbf{v}=v\mathbf{\hat{z}}$, so the linearised perturbed MHD equations in a 1D system are
\begin{eqnarray}
&&\frac{\partial\rho_1}{\partial t}+\rho_0\frac{\partial v_1}{\partial z}=0,\label{Eq:linearised-cont.}\\
&&\rho_0\frac{\partial v_1}{\partial t}=-\frac{\partial p_1}{\partial z},\label{Eq:linearised-motion}\\
&&\frac{R}{\tilde\mu}\left[\rho_{1}\frac{\md{T}_{0}}{\md{t}}+\rho_{0}\frac{\partial{T}_{1}}{\partial{t}}+({\gamma-1})\rho_{0}T_{0}
\frac{\partial{v}_{1}}{\partial z}\right]=(\gamma-1)\kappa_0T_0^{5/2}\frac{\partial^2T_1}{\partial z^2},\label{Eq:linearised-energy}\\
&&p_1=\frac{R}{\tilde\mu}\left\{\rho_0T_1+T_0\rho_1\right\}\label{Eq:linearised-gas_law}.
\end{eqnarray}
Here $v_1\equiv v_{1z}$ is the longitudinal velocity perturbation. It is clear that there are no terms exhibiting the magnetic field  but the standing waves are still guided by the magnetic field.
Non-dimensionalisation will be used to simplify the 1D governing Equations $(\ref{Eq:linearised-cont.})-(\ref{Eq:linearised-gas_law})$. The dimensionless quantities are introduced as
\begin{eqnarray}
&&\tilde{t}=\frac{t}{P},\quad\tilde{z}=\frac{z}{L},\quad c_{si}=\frac{L}{P},\quad\tilde{p}_{0}=\frac{p_{0}}{p_{0i}},\quad\tilde{T}_{0}=\frac{T_{0}}{T_{0i}},\nonumber
\\&& c_{si}^2=\frac{\gamma{p}_{0i}}{\rho_{0}},
\quad\tilde\rho_{1}=\frac{\rho_{1}}{\rho_{0}},\quad\tilde{p}_{1}=\frac{p_{1}}{p_{0i}},\quad \tilde{T}_{1}=\frac{T_{1}}{T_{0i}}, \quad \tilde{v}_{1}=\frac{{v}_{1}}{c_{si}},
\end{eqnarray}
where the subscript $i$ represents the value of the quantity at $t=0$, $L$ is the loop length, and $c_{si}$ is the initial sound speed.

Now, we aim to find the governing equation for the perturbed velocity, which leads to reveal the behaviour of the standing magneto--acoustic mode subject to initial conditions generated in a hot coronal loop. Using the continuity and ideal gas law equations, Equation ($\ref{Eq:linearised-energy}$) in terms of dimensionless variables, removing the tilde for the sake of simplicity, takes the following form
\begin{equation}
\frac{\partial^2{v}_{1}}{\partial t^2}-T_0\frac{\partial^2{v}_{1}}{\partial z^2} =\frac{-\sigma}{\gamma}T^{5/2}_0\frac{\partial^3T_1}{\partial z^3}, \qquad \sigma=\frac{(\gamma-1)\tilde{\mu}\kappa_0\,T^{5/2}_{0i}}{RL\sqrt{\gamma\, p_{0i}\,\rho_0}},\label{Eq:nondimen-energy}
\end{equation}
where $\sigma$ is defined as thermal ratio and found to be a small quantity under typical coronal conditions (see, {\it e.g}. \opencite{Moortel03}) demonstrated here as
\begin{equation}
\left\{
\begin{array}{ll}
T_0=6\times10^5-5\times10^6\;\textmd{K},\\
\rho_0=1.67\times10^{-12}\; \textmd{kg m}^{-3},\\
\kappa_{\|}=10^{-11}\,T_0^{5/2}\; \textmd{W m}^{-1}\; \textmd{deg}^{-1},\\
\tilde{\mu}=0.6,\\
R=8.3\times10^3\;\textmd{m}^2\;\textmd{s}^{-2}\;\textmd{deg}^{-1},\\
\gamma=5/3,\\
L=10^8\; \textmd{m},\\
\end{array}
\right.
\end{equation}
giving characteristic values of $\sigma\in[0.0068,0.48]$ for $T\in[6\times10^5,5\times10^6]$ in K.

Equation ($\ref{Eq:nondimen-energy}$) with the aid of Equations ($\ref{Eq:back-motion}$), ($\ref{Eq:back-gas_law}$) and ($\ref{Eq:linearised-gas_law}$) will be
\begin{equation}
\frac{1}{T^{7/2}_0}\left(\frac{\partial^2{v}_{1}}{\partial t^2}-T_0\frac{\partial^2{v}_{1}}{\partial z^2}\right) =\frac{-\sigma}{\gamma}\frac{\partial^3}{\partial z^3}\left[\frac{p_1}{T_0}-\rho_1\right].\label{Eq:modified-nondimen-energy}
\end{equation}
Differentiating with respect to time and, using Equations ($\ref{Eq:linearised-cont.}$) and ($\ref{Eq:linearised-motion}$), we obtain
\begin{equation}
\frac{\partial}{\partial t}\left[\frac{1}{T^{7/2}_0}\left(\frac{\partial^2{v}_{1}}{\partial t^2}-T_0\frac{\partial^2{v}_{1}}{\partial z^2}\right)\right]=\frac{\sigma}{\gamma}\left[\frac{\gamma}{T_0}\frac{\partial^4v_1}{\partial t^2 \partial z^2}+\gamma\frac{\md}{\md t}(\frac{1}{T_0})\frac{\partial^3v_1}{\partial t \partial z^2}-\frac{\partial^4v_1}{\partial z^4}\right],
\end{equation}
which is the governing equation and can be recast to a convenient form for further analysis:
\begin{equation}
\frac{\partial}{\partial t}\left(\frac{\partial^2{v}_{1}}{\partial t^2}-T_0\frac{\partial^2{v}_{1}}{\partial z^2}\right)=\frac{7}{2}\frac{1}{T_0}\frac{\md T_0}{\md t}\left(\frac{\partial^2{v}_{1}}{\partial t^2}-T_0\frac{\partial^2{v}_{1}}{\partial z^2}\right)-\sigma T^{3/2}_0\frac{\md T_0}{\md t}\frac{\partial^3v_1}{\partial t \partial z^2}+\sigma T^{5/2}_0\frac{\partial^2}{\partial z^2}\left(\frac{\partial^2{v}_{1}}{\partial t^2}-\frac{T_0}{\gamma}\frac{\partial^2{v}_{1}}{\partial z^2}\right).\label{Eq:modified-governing}
\end{equation}
There are three different cases with interest for the present context one can recover from Equation ($\ref{Eq:modified-governing}$):\\
\begin{description}
\item[Case I.] In the absence of the thermal conduction, $\sigma$, and the unspecified thermodynamic source in the energy equation, $\mathcal{L}\propto~\md T_0/\md t$, the governing equation reduces to
\begin{equation}
\frac{\partial^2{v}_{1}}{\partial t^2}-c_s^2\frac{\partial^2{v}_{1}}{\partial z^2}=0,\qquad c_s=\sqrt{T_0}=\textrm{const.},\label{Eq:wave}
\end{equation}
which has the solution
\begin{equation}
v_1(z,t)=\alpha\cos(\pi z)\cos(\pi c_s t),
\end{equation}
under the boundary conditions representing a line-tied flux tube perturbed as a semi-circular initial motion proportional to its fundamental mode
\begin{equation}
v_1(t,\pm\frac{1}{2})=0,\quad v_1(0,z)=\alpha\cos(\pi z),\quad \frac{\partial v_1}{\partial t}(0,z)=0,\label{Eq:boundary conditions}
\end{equation}
where $\alpha$ is the initial amplitude of the standing wave at $t=0$. A more general, {\it e.g.} broad-band, perturbation would, of course, give the solution in the mathematical form of a Fourier-series.
\item[Case II.] In the case of no thermal conduction, {\it i.e.} $\sigma=0$, the effect of unspecified thermodynamic source in the system will be found by solving the following equation
\begin{equation}
\frac{\partial^2{v}_{1}}{\partial t^2}-c_s^2\frac{\partial^2{v}_{1}}{\partial z^2}=0,\qquad c_s(t)=\sqrt{T_0(t)}\neq\textrm{const.}, \label{Eq:governing-no thermal_cond}
\end{equation}
which is {\it formally} exactly Equation ($\ref{Eq:wave}$) but with variable background temperature, $T_0=T_0(t)$. In spite of the absence of $\sigma$, the coefficient of the bracket in the first term in the right-hand-side of Equation ($\ref{Eq:modified-governing}$) is originally derived from thermal conduction term as seen in Equation ($\ref{Eq:modified-nondimen-energy}$). Therefore, this term is not added to Equation ($\ref{Eq:governing-no thermal_cond}$).
\item[Case III.] The effect of thermal conduction on the behaviour of the standing acoustic waves in the presence of an unspecified thermodynamic source will be investigated by solving Equation ($\ref{Eq:modified-governing}$).
\end{description}
\section{Analytical Solutions}
Our goal now is to find an analytic solution to the governing Equation ($\ref{Eq:governing-no thermal_cond}$) first in case II and next to Equation ($\ref{Eq:modified-governing}$) in case III. Let us point out that Equations ($\ref{Eq:modified-governing}$) and ($\ref{Eq:governing-no thermal_cond}$) have derivatives multiplied by small factors $\sigma$ and $\epsilon$ so this enables the use of the WKB theory to obtain an approximate solution, where the WKB estimates give good approximations when the smaller value of factor used.

\bigskip
{\bf Case II.} Let us first start to solve Equation ($\ref{Eq:governing-no thermal_cond}$) to establish the behaviour of standing magneto--acoustic waves under the influence of an unspecified thermodynamic source in a cooling background state. Assuming that $t_1=\epsilon t$ which is defined as a slow timescale, meaning the cooling timescale is (much) longer than the period of the oscillations, Equation ($\ref{Eq:governing-no thermal_cond}$) will reduce to
\begin{equation}
\frac{\partial^2{v}_{1}}{\partial t_1^2}-\frac{c_s^2}{\epsilon^2}\frac{\partial^2 v_1}{\partial z^2}=0.\label{Eq:governing-no thermal_cond-WKB}
\end{equation}
In line with applying the WKB estimates let the perturbed velocity variable have the form
\begin{equation}
v_1(z,t_1)=Q(z,t_1)\exp\left(\frac{i}{\epsilon}\Theta(t_1)\right).\label{Eq:wkb1}
\end{equation}
The amplitude $Q$ can be expanded  in the power series as follows
\begin{equation}
Q(z,t_1)=Q_0+\epsilon\, Q_1+\cdots.\label{Eq:power-series}
\end{equation}
Substituting Equations ($\ref{Eq:wkb1}$) and ($\ref{Eq:power-series}$) into Equation ($\ref{Eq:governing-no thermal_cond-WKB}$), and taking terms of order $\epsilon^{-3}$ we obtain at leading order
\begin{equation}
\frac{\partial^2Q_0}{\partial z^2}+\frac{\omega^2}{c_s^2}Q_0=0,\label{Eq:highest-order}
\end{equation}
where $\omega=\md\Theta/\md t_1$. The boundary conditions ($\ref{Eq:boundary conditions}$) applicable to the function $Q_0$ is
\begin{equation}
Q_0=0 \qquad \textrm{at}\quad z=\pm\frac{1}{2}.\label{Eq:condition-highest-order}
\end{equation}
Equations ($\ref{Eq:highest-order}$) and ($\ref{Eq:condition-highest-order}$) represent a boundary-value problem that determines the frequency of the standing longitudinal (acoustic) mode in a cooling plasma with a varying temperature as function of time. The general solution physically acceptable to this boundary value problem has the form
\begin{equation}
Q_0(z,t_1)=\sum_{n=0}^{\infty} A_n(t_1)\cos\left((2n+1)\pi z\right), \qquad \omega_n=(2n+1)\pi c_s,\;n=0,1,2,\cdots.\label{Eq:Standing_Unspecified_therm-source}
\end{equation}
Then, collecting terms of order $\epsilon^{-2}$, we obtain the equation determining the amplitude
\begin{equation}
\frac{\partial^2Q_1}{\partial z^2}+\frac{\omega^2}{c_s^2}Q_1=\frac{i}{c_s^2}\left[\frac{\mathrm{d}\omega}{\mathrm{d} t_1}Q_0+2\omega\frac{\partial Q_0}{\partial t_1}\right].\label{Eq:next-order}
\end{equation}
It follows from Equation ($\ref{Eq:boundary conditions}$) that $Q_1$ satisfies the boundary conditions
\begin{equation}
Q_1=0 \qquad \textrm{at}\quad z=\pm\frac{1}{2}.\label{Eq:condition-next-order}
\end{equation}
The boundary-value problem, Equations ($\ref{Eq:next-order}$) and ($\ref{Eq:condition-next-order}$), determining $Q_1$ is a Sturm-Liouville problem and has a solution only when the right-hand-side of Equation ($\ref{Eq:next-order}$) satisfies the compatibility condition, which is the condition that it is orthogonal to $Q_0$ (see, \opencite{Ruderman11}). This condition can be obtained by multiplying Equation ($\ref{Eq:next-order}$) by $Q_0$, integrating with respect to $z$ from $-1/2$ to $1/2$ and using the boundary conditions ($\ref{Eq:condition-next-order}$). The compatibility condition is eventually written as
\begin{equation}
\int_{-1/2}^{1/2}\frac{i}{c_s^2}\left[\frac{\mathrm{d}\omega}{\mathrm{d} t_1}Q_0^2+2\omega Q_0\frac{\partial Q_0}{\partial t_1}\right]\mathrm{d}z=0,
\end{equation}
which gives the amplitude of the standing wave in the following form
\begin{equation}
A_n(t_1)=A_n(0)\exp(\frac{t_1}{4})=A_n(0)\sqrt{\frac{c_s(0)}{c_s(t_1)}}.\label{Eq:Amp_Unspecified_therm-source}
\end{equation}
The value of constant $A_n(0)$ can be found from Equation $(\ref{Eq:Standing_Unspecified_therm-source})$ at $t_1=0$, which is in the form of Fourier cosine series, using the boundary condition $(\ref{Eq:boundary conditions})$. Then, the solution ($\ref{Eq:Amp_Unspecified_therm-source}$), in scaled ({\it i.e.} physical) variables, takes the form
\begin{equation}
A_n(t)=\alpha\,\sqrt{\frac{c_s(0)}{c_s(t)}}.\label{Eq:modified-Amp_Unspecified_therm-source}
\end{equation}
This result describes the variation of a time-dependent amplitude of longitudinal standing waves in a dynamically cooling magnetic flux tube. Apparently, Equation ($\ref{Eq:modified-Amp_Unspecified_therm-source}$) indicates that the cooling (or heating) leads to an amplification (decrease) of loop oscillation.

\bigskip
{\bf Case III.} Next, the behaviour of the standing wave in a system dominated by thermal conduction will be obtained by solving the governing Equation ($\ref{Eq:modified-governing}$). Similarly, the WKB theory will be used to find the solution of Equation  ($\ref{Eq:modified-governing}$). Let us now introduce two slow variables $t_1=\epsilon t$ and $\sigma=\epsilon\tilde{\sigma}$, so Equation ($\ref{Eq:modified-governing}$) will be
\begin{equation}
\frac{\partial^3{v}_{1}}{\partial t_1^3}+\frac{7}{2}\frac{\partial^2{v}_{1}}{\partial t_1^2}-\frac{c_s^2}{\epsilon^2}\frac{\partial^3{v}_{1}}{\partial t_1\partial z^2} -\tilde{\sigma}c_s^5\frac{\partial^3v_1}{\partial t_1 \partial z^2} -\tilde{\sigma}c_s^5\frac{\partial^4{v}_{1}}{\partial t_1^2\partial z^2}-\frac{5}{2}\frac{c_s^2}{\epsilon^2}\frac{\partial^2 v_1}{\partial z^2}+\frac{\tilde{\sigma}}{\epsilon^2}\frac{c_s^7}{\gamma}\frac{\partial^4{v}_{1}}{\partial z^4}=0,\label{Eq:governing-thermal_cond-WKB}
\end{equation}
and the perturbed velocity by the new scaled variables and the WKB approximation is given by Equation ($\ref{Eq:wkb1}$).
Substituting Equations ($\ref{Eq:wkb1}$) and ($\ref{Eq:power-series}$) into Equation ($\ref{Eq:governing-thermal_cond-WKB}$), and taking the highest order terms in $\epsilon$, which is again $\epsilon^{-3}$, we obtain
\begin{equation}
\frac{\partial^2Q_0}{\partial z^2}+\frac{\omega^2}{c_s^2}Q_0=0, \label{Eq:governing_highest-order}
\end{equation}
with the boundary conditions
\begin{equation}
Q_0=0 \qquad \textrm{at}\quad z=\pm\frac{1}{2}.
\end{equation}
The solution to this boundary-value problem is given by
\begin{equation}
Q_0(z,t_1)=\sum_{n=0}^{\infty} B_n(t_1)\cos\left((2n+1)\pi z\right), \qquad \omega_n=(2n+1)\pi c_s,\;n=0,1,2,\cdots,\label{Eq:sol-highest-order}
\end{equation}
where $B_n(t_1)$ stands for the amplitude of the longitudinal standing modes and it will be found by taking the second highest order terms in $\epsilon$ for Equation ($\ref{Eq:governing-thermal_cond-WKB}$) and then using the properties of Sturm-Liouville problems as follows.

The next largest order terms in $\epsilon$, of order $\epsilon^{-2}$, give the following equation
\begin{equation}
\frac{\partial^2Q_1}{\partial z^2}+\frac{\omega^2}{c_s^2}Q_1=\frac{i}{\omega c_s^2}\left[(\frac{7}{2}\omega^2+3\omega\frac{\md\omega}{\md t_1})Q_0+3\omega^2\frac{\partial Q_0}{\partial t_1}+c_s^2\frac{\partial^3Q_0}{\partial t_1\partial z^2}+(\frac{5}{2}c_s^2-\tilde{\sigma}\omega^2c_s^5)\frac{\partial^2Q_0}{\partial z^2}-\tilde{\sigma}\frac{c_s^7}{\gamma}\frac{\partial^4{Q}_{0}}{\partial z^4}\right].\label{Eq:governing_next-highest-order}
\end{equation}
It yields from Equation ($\ref{Eq:boundary conditions}$) that $Q_1$ satisfies the boundary conditions
\begin{equation}
Q_1=0 \qquad \textrm{at}\quad z=\pm\frac{1}{2}.\label{Eq:governing_condition-next-order}
\end{equation}
Analogous to Equations ($\ref{Eq:next-order}$) and ($\ref{Eq:condition-next-order}$), this boundary-value problem, Equations ($\ref{Eq:governing_next-highest-order}$) and ($\ref{Eq:governing_condition-next-order}$), has a solution only when the right-hand side of Equation ($\ref{Eq:governing_next-highest-order}$) satisfies the compatibility condition. Consequently,
\begin{eqnarray}
\int_{-1/2}^{1/2}\frac{i}{\omega c_s^2}\left[(\frac{7}{2}\omega^2+3\omega\frac{\md\omega}{\md t_1})Q_0^2+3\omega^2Q_0\frac{\partial Q_0}{\partial t_1}+c_s^2Q_0\frac{\partial^3Q_0}{\partial t_1\partial z^2}+(\frac{5}{2}c_s^2-\tilde{\sigma}\omega^2c_s^5)Q_0\frac{\partial^2Q_0}{\partial z^2}\right.\nonumber\\-\left.\tilde{\sigma}\frac{c_s^7}{\gamma}Q_0\frac{\partial^4{Q}_{0}}{\partial z^4}\right]\md z=0.\label{Eq:compatibility_cond}
\end{eqnarray}
Substituting ($\ref{Eq:sol-highest-order}$) into ($\ref{Eq:compatibility_cond}$), we obtain the amplitude of standing wave
\begin{equation}
B_n(t_1)=B_n(0)\exp\left(\frac{1}{4}t_1+\frac{\tilde{\sigma}}{5}(\frac{\gamma-1}{\gamma})(2n+1)^2\pi^2(c_s^5-1)\right),
\end{equation}
which can be re-written in the scaled variables as
\begin{equation}
B_n(t)=B_n(0)\exp\left(\frac{\epsilon}{4}t+\frac{\sigma}{5\epsilon}(\frac{\gamma-1}{\gamma})(2n+1)^2\pi^2(c_s^5-1)\right).\label{Eq:Amp_thermal-conduction}
\end{equation}
Now, Equation ($\ref{Eq:sol-highest-order}$), which represents a Fourier cosine series, with the boundary conditions ($\ref{Eq:boundary conditions}$) will be applied to obtain the value of constants $B_n(0)$. Hence, the solution ($\ref{Eq:Amp_thermal-conduction}$) is
\begin{equation}
B_n(t)=\alpha\exp\left(\frac{\epsilon}{4}t+\frac{\sigma}{5\epsilon}(\frac{\gamma-1}{\gamma})(2n+1)^2\pi^2(c_s^5-1)\right),\label{Eq:modified-Amp_thermal-conduction}
\end{equation}
which reveals the temporal evolution of longitudinal standing-mode amplitude due to thermal conduction in a temporally variable (cooling or heating) background plasma.
Note that, in the limit $\sigma\rightarrow0$, {\it  i.e.} if there is no thermal conduction, Equation ($\ref{Eq:modified-Amp_thermal-conduction}$) reduces to Equation ($\ref{Eq:modified-Amp_Unspecified_therm-source}$) which represents the amplitude variation of standing waves generated in a model of merely cooling plasma without non-ideal effects as, {\it e.g.}, thermal conduction.

\section{Numerical Evaluations}
\inlinecite{Morton09d} and \inlinecite{Erdelyi11} have shown that the WKB approximation estimates accurately the solutions to the frequency and amplitude variations in time and space for waves supported by oscillating magnetic loops due to plasma cooling by radiation and/or thermal conduction, respectively.  The obtained approximate solutions by the WKB theory can be motivated using numerical evaluations to demonstrate a clear view of the behaviour of MHD waves which is analytically found.

\begin{figure}[!ht]
\centering
\includegraphics[height=0.45\textheight,width=0.6\textwidth]{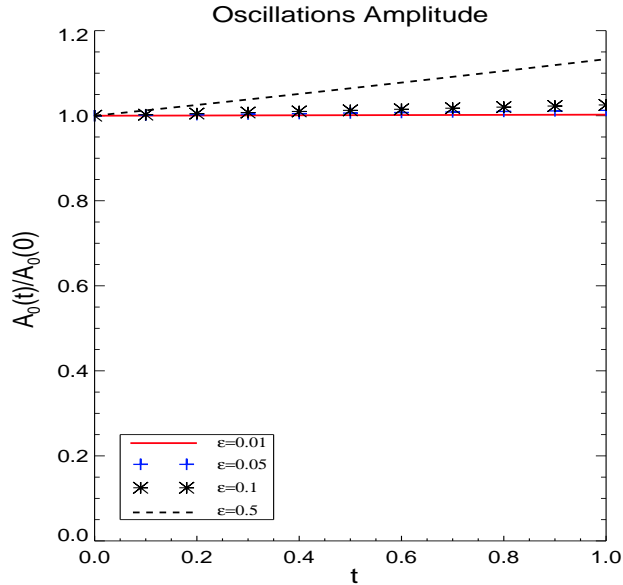}
\vspace{-0.8cm}
\caption{The amplitude of the standing wave with different values of $\epsilon$ $(0.01, 0.05, 0.1, 0.5)$ representing the ratio of period to the cooling time.}\label{Amp_unspecified source}
\end{figure}

The amplitude of the longitudinal (acoustic) standing waves is plotted after calculating the variables using standard coronal values. Figure \ref{Amp_unspecified source} shows the variations of the amplitude due to cooling (or heating) mechanism. This graph exhibits an amplification for the wave amplitude caused by the cooling of the background where increasing $\epsilon$ (the ratio of the period of oscillation to the cooling time scale) increases the amplitude. It should be mentioned that this result is in agreement with that reached by \inlinecite{Erdelyi11} who found that the efficiency of damping is reduced by the cooling background plasma, where Figure \ref{Amp_unspecified source} displays no damping for standing oscillation because the model is purely dominated by the cooling.

\begin{figure}[!ht]
\centering
\hspace{-0.8 cm}
\subfloat{\label{Amp-damping2a}\includegraphics[height=0.4\textheight, width=0.375\textwidth]{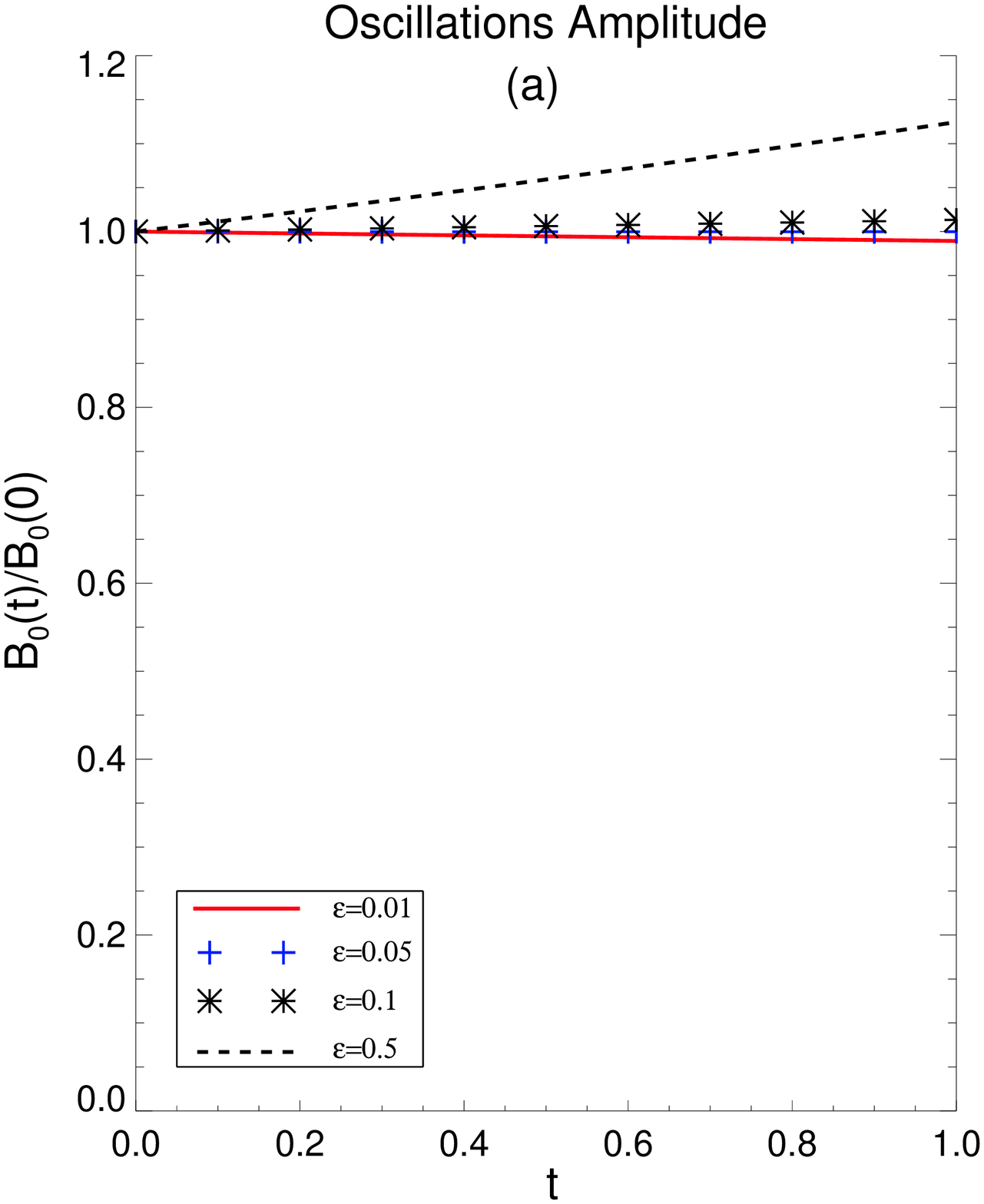}}
\hspace{-0.9 cm}
\subfloat{\label{Amp-damping2b}\includegraphics[height=0.4\textheight, width=0.375\textwidth]{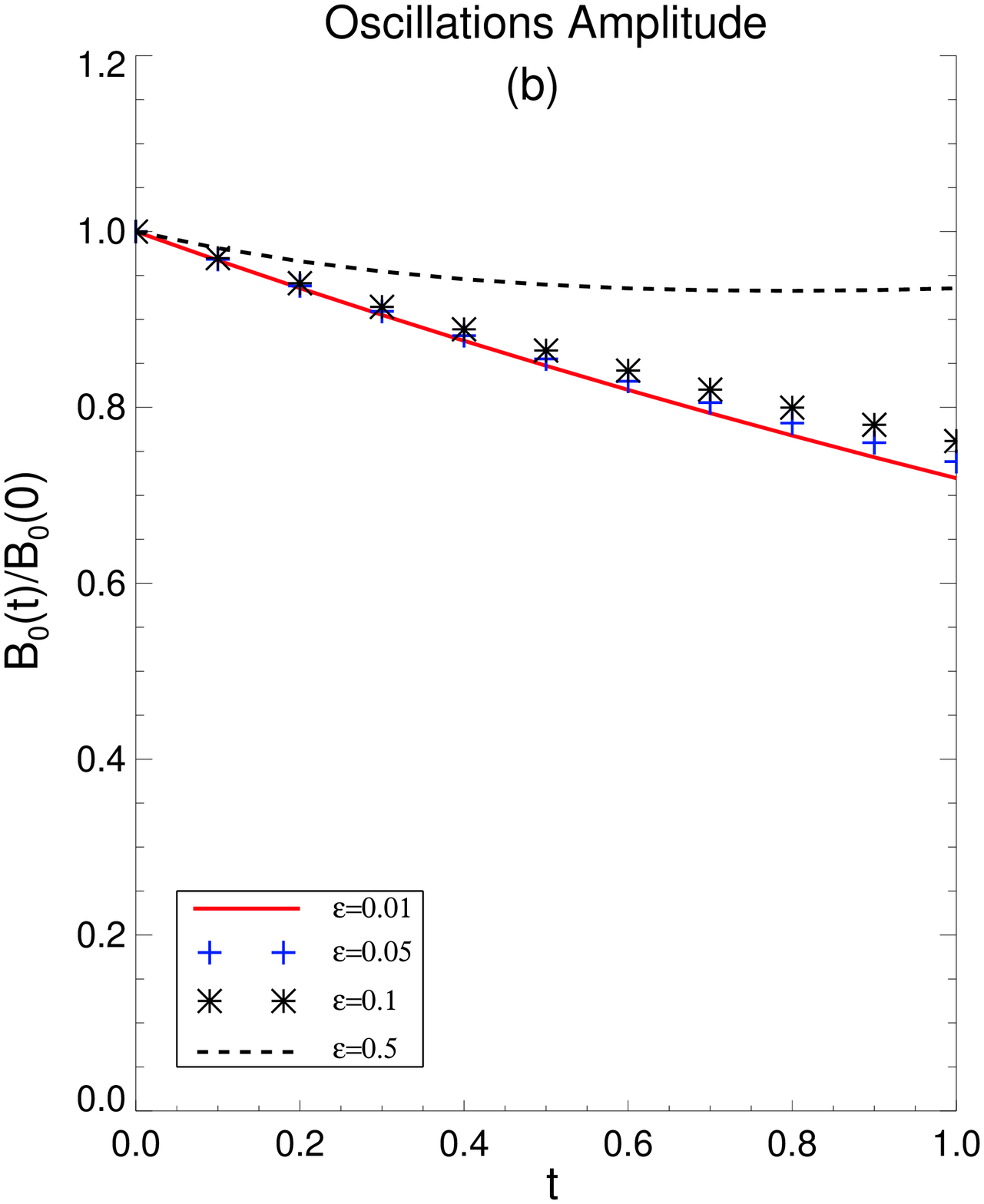}}
\hspace{-0.9 cm}
\subfloat{\label{Amp-damping2c}\includegraphics[height=0.4\textheight, width=0.375\textwidth]{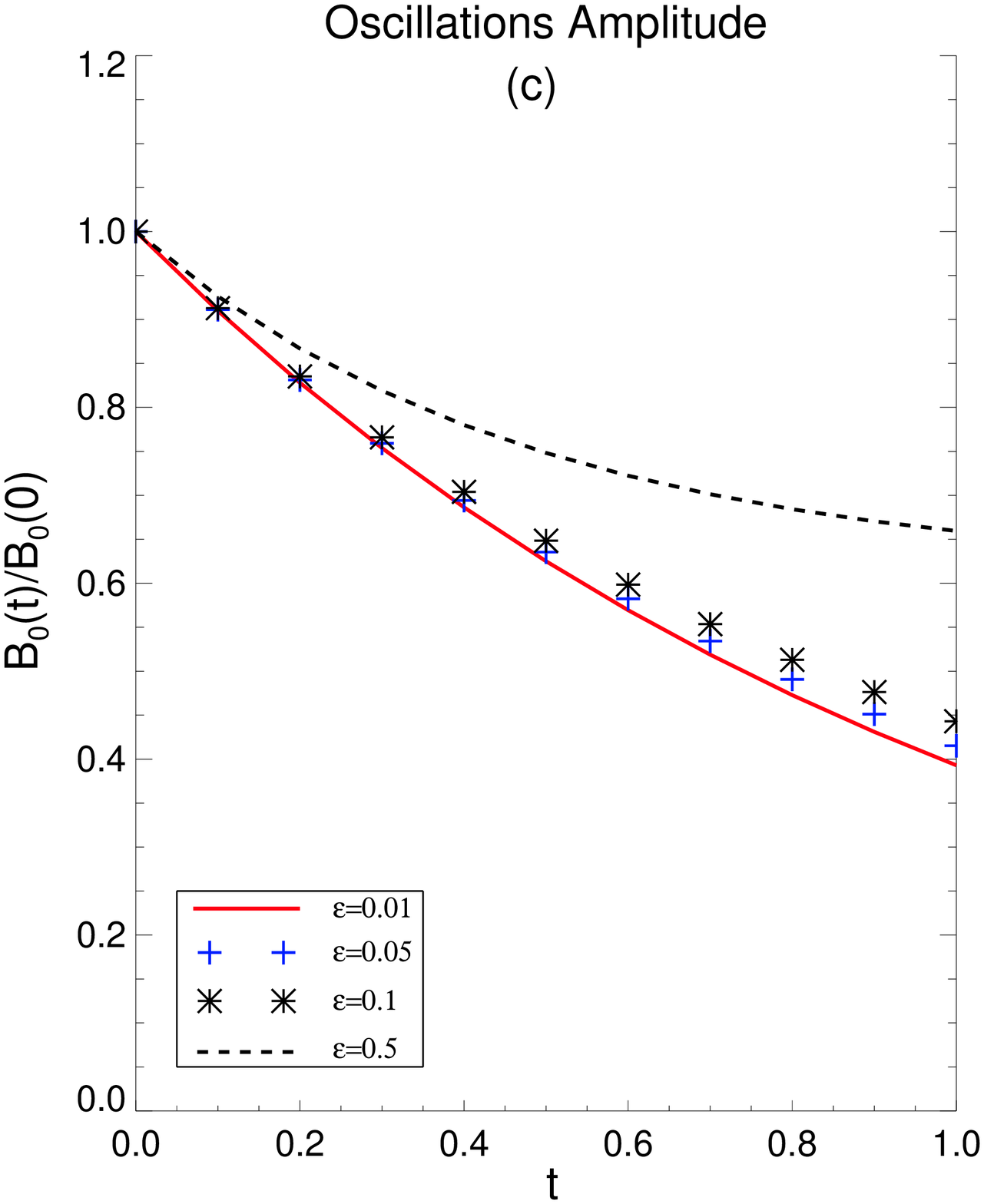}}
\hspace{-0.8cm}
\vspace{-1 cm}
\caption{The amplitude of the standing wave with different values of $\epsilon$ $(0.01, 0.05, 0.1, 0.5)$ representing the ratio of period to the cooling time and specific value of $\sigma$, {\it i.e.} the value of thermal ratio. (a) $\sigma=0.0068$~($T=600000$~K), (b) $\sigma=0.17$ ($T=3$~MK),  (c) $\sigma=0.48$ ($T=5$~MK).}\label{Amp-damping}
\end{figure}

Next, in Figure \ref{Amp-damping} we demonstrate how the amplitude of standing longitudinal (acoustic) waves for a range of values of $\epsilon$ and as function of the value of thermal ratio $\sigma$ is changing in different temperature regions. It is found that the variation of $\epsilon$ leads to a considerable change in the rate of damping of both cool and hot loops. Figure \ref{Amp-damping2a} illustrates the trend of amplitude of the EUV (cool) loops such that the oscillation amplitude decreases slowly in regions of temperature $600,000$~K. The decay of the wave amplitude increases slightly with time in hot corona, for instance loops of temperature 3 MK and 5 MK as depicted in Figure \ref{Amp-damping2b} and \ref{Amp-damping2c}, respectively. The strength of damping of hot ({\it e.g.} SXT/XRT) loops is found to be much stronger and shows little change for the smallest values of $\epsilon$, {\it i.e.} $0.01-0.1$, whereas large enough values of $\epsilon$, {\it i.e.} in the range $0.1<\epsilon\le0.5$, cause a rapid reduction in the rate of damping of the wave amplitude. Moreover, it is obvious from Equation ($\ref{Eq:Amp_thermal-conduction}$) that the first term in the exponential function is the same as in Equation ($\ref{Eq:Amp_Unspecified_therm-source}$) were dominated by the cooling. This indicates that the emergence of cooling by thermal conduction in the system of hot coronal loops decreases the rate of damping as exhibited in Figure \ref{Amp-damping}.

\begin{figure}[h]
\centering
\includegraphics[height=0.45\textheight,width=0.6\textwidth]{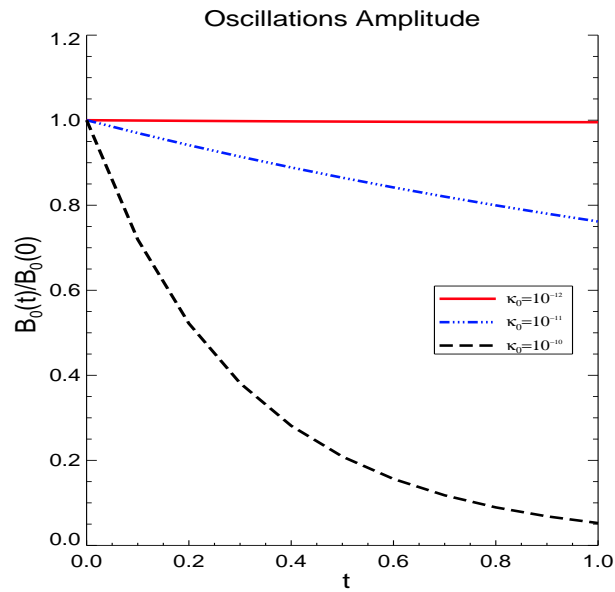}
\vspace{-0.8cm}
\caption{The amplitude of the standing wave with different values of the thermal-conduction coefficient, $\kappa_0$=$(10^{-10}, 10^{-11}, 10^{-12})$ and specific value of the ratio of period to the cooling time, $\epsilon=0.1$ where $T=3$~MK.}\label{thermal_conduction}
\end{figure}

In Figure $\ref{thermal_conduction}$ we present the effect of varying the magnitude of thermal conduction coefficient, $\kappa$, on the rate of damping of the standing acoustic wave. Typical values for the coefficient of the thermal conductivity $\kappa=[10^{-10},10^{-11},10^{-12}]T^{5/2}$ are taken to shed light on the influence of thermal conduction on hot corona, $T=3$ MK, and the characteristic value of $\epsilon=0.1$ is assumed (see, \opencite{Priest}). It is found that the rate of damping is changing rather rapidly with altering $\kappa$ by just an order of magnitude where increasing the value of $\kappa$ gives rise to a strong decline in the amplitude of the standing slow mode.

The significance of the obtained results is to be comparable to observations. We only present here a quantitative comparison with the properties of hot loop oscillations observed by SUMER (see, \opencite{Wang03a}), where the standing slow-mode waves are detected only in the region of temperature $\ge6$ MK. The periods of oscillations are $7-31$ minutes. The slow standing wave has seen to be strongly damped with characteristic decay times $5.7-36.8$ minutes mainly likely due to thermal conduction. The typical length of coronal loops is around 230 Mm.

In our work, we found that hot loop oscillations experience a strong damping due to thermal conduction which might be comparable with the observed damping provided the value of $\epsilon$ is small enough as shown in Figure \ref{Amp-damping2c}. Further to this, Figure \ref{thermal_conduction} exhibits  that the large value of thermal conduction coefficient leads to a rapid damping which is likely applicable for the observed damping of standing acoustic modes as discussed by \inlinecite{Ofmanwang} and \inlinecite{Mendoza04}.
\section{Discussion and Conclusion}
In this work, we have investigated the influence of a cooling background on the standing magneto-acoustic waves generated in a uniformly magnetised plasma. Thermal conduction is assumed to be the dominant mechanism of cooling background plasma. The background temperature is allowed to change as a function of time and to decay exponentially with characteristic cooling times typical for coronal loops. The magnetic field is assumed to be constant and in the $z$ direction which may be a suitable model for loops with large aspect ratio. A theoretical model of 1D geometry describing the coronal loop is applied. A time-dependant governing equation is derived by perturbing the background plasma on a time-scale greater than the period of the oscillation. Three different cases were considered: (I) absence of thermal conduction $\sigma$ and unspecified cooling or heating mechanism $\mathcal{L}$, (II) presence of unknown thermodynamic source $\mathcal{L}$ only, {\it i.e.} $\sigma=0$, (III) the influence of thermal conduction $\sigma$ combined with the unknown mechanism $\mathcal{L}$.

The WKB theory is used to find the analytical solution of the governing equation in case II and III where the governing equation in case I is solved by a direct method and gives the undamped standing wave, {\it i.e.} sound speed is constant.  An approximate solution that describes a time-dependant amplitude of the standing acoustic mode is obtained with the aid of the properties of a Sturm-Liouville problem. The analytically derived solutions are exhibited numerically to give much illustration to the behaviour of MHD slow waves.

In the second case, the individual influence of cooling background plasma on hot-loop oscillation is found to cause an amplification to the amplitude of the longitudinal standing wave. It is noted that the rate of amplification varies according to the change in the ratio of the oscillatory period to the cooling time scale $\epsilon$, where increasing $\epsilon$ increases the amplitude of MHD wave.

In the third case, which is of our interest, the temporally dependant amplitude is found to undergo a strong damping due to the cooling of the background plasma by thermal conduction in the hot corona. Further to this, we note that the presence of cooling in a model of hot coronal loop decreases the efficiency of damping. The variation of the ratio of the period of oscillation to the cooling time scale, $\epsilon$, plays an effective role on changing the rate of damping of oscillating hot coronal loops, causing a fast decline in the decay degree of oscillation amplitude once the value of this ratio is so large.

The obtained results indicate that this investigation contributes with the previous studies (\opencite{Morton09b}, \citeyear{Morton09c}; \opencite{Morton09d}; \opencite{Erdelyi11}) on demonstrating that the temporal evolution of coronal plasma due to the dissipative process, {\it i.e.} the cooling of the background plasma due to radiation/thermal conduction, has a great influence on the coronal oscillations. In the modelling of solar coronal loop, the temporal and spatial dependant dynamic background plasma is necessary to be considered to understand the properties of observed MHD waves.

\bigskip

\begin{Ack}
The authors would like to thank M.S. Ruderman and R.J. Morton for useful
discussion. R.E. acknowledges M. K\'{e}ray for patient
encouragement. The authors are also grateful to NSF, Hungary (OTKA,
Ref. No. K83133), Science and Technology Facilities Council (STFC),
UK and Ministry of Higher Education, Oman for the financial support.
\end{Ack}


\bibliographystyle{spr-mp-sola}

\end{article}
\end{document}